# A review of artificial intelligence methods combined with Raman spectroscopy to identify the composition of substances

Liangrui Pan[1], Peng Zhang[1], Chalongrat Daengngam[2] and Mitchai Chongcheawchamnan[1]

[1]Department of Electrical Engineering, Faculty of Engineering, Prince of Songkla University, Songkhla, Thailand
[2]Division of Physical Science, Faculty of Science, Prince of Songkla University, Songkhla, Thailand

**Abstract**
In general, most of the substances in nature exist in mixtures, and the noninvasive identification of mixture composition with high speed and accuracy remains a difficult task. However, the development of Raman spectroscopy, machine learning, and deep learning techniques have paved the way for achieving efficient analytical tools capable of identifying mixture components, making an apparent breakthrough in the identification of mixtures beyond the traditional chemical analysis methods. This article summarizes the work of Raman spectroscopy in identifying the composition of substances as well as provides detailed reviews on the preprocessing process of Raman spectroscopy, the analysis methods and applications of artificial intelligence. This review summarizes the work of Raman spectroscopy in identifying the composition of substances and reviews the preprocessing process of Raman spectroscopy, the analysis methods and applications of artificial intelligence. Finally, the advantages and disadvantages and development prospects of Raman spectroscopy are discussed in detail.
Keywords: preprocessing, Raman spectroscopy, artificial intelligence, composition

## 1 Introduction

Material-based component analysis and identification, chemical analysis, high performance liquid chromatography (HPLC), infrared spectroscopy and Raman spectroscopy showed promising results in the experiment. However, the chemical analysis method is long, time-consuming and suitable for constant grouping. HPLC is expensive, requires various packing columns, has a small capacity, and is challenging to analyze biological macromolecules and inorganic ions. The mobile phase is mostly consumed and toxic. Infrared spectroscopy has high sensitivity and low cost and has certain limitations in use. Raman spectroscopy has the advantages of faster analysis speed, high sensitivity, low cost, and simple operation compared with these methods.

Raman spectroscopy is one of the spectroscopy techniques based on molecular vibration that has been widely used in chemical identification in recent decades. Raman scattering process is associated with unique vibrational modes of molecules. Thus, the Raman spectrum contains useful information about molecular bonds and structures, possibly identifying chemical components in a substance. Therefore, this technique is widely used in molecular structure identification, industrial process control, planetary science exploration, life science research and field geological expeditions [1]–[4]. Like most spectroscopic techniques, Raman spectroscopy requires advanced data processing techniques to extract valuable raw spectra information. The process involves cosmic ray artefacts removal, spectrum smoothing and baseline correction [5]–[7]. Linear regression and multivariate data analysis methods have been usually used to achieve high accuracy, sensitivity, and selective identification of vibrational fingerprints to analyze complex substance signals. Although supervised learning algorithms are currently the most useful classification methods, raw Raman spectra need to be preprocessed before using them.

In general, an unprocessed Raman spectrum consists of chemical information, baselines and random noise [8]. The presence of baselines and random noise can negatively affect the results of the qualitative analysis of substances. Therefore, raw Raman spectrum cannot be directly used for identifying. A baseline noise that usually occurres as a broader spectrum unrelated to the sample composition should be removed after spectrum acquisition to minimize adverse effects. Standard processing methods need to point out the two ends of the Raman spectrum signal peak and then use the piecewise linear approximation method to fit the curve as the baseline[9]. However, standard processing methods are not very useful, and their accuracy depends on human skills. In recent years, researchers have proposed several algorithms for automatic baseline correction, and these methods will significantly improve the accuracy of qualitative analysis of substances. The typical baseline correction methods are divided into two main categories, polynomial baseline modeling and least-square technique [10], [11].

Several baseline correction algorithms were proposed based on the polynomial fitting method [12]–[15]. Mauro et al. proposed a polynomial fitting estimation of the baseline to eliminate the baseline drift in complex environmental samples [12]. Feng et al. introduced an improved iterative polynomial fitting with the automatic threshold for baseline correction [14]. Julia et al. proposed a correction algorithm based on polynomial fitting calculated from the reference spectrum matrix (Polyfit-RSM method) [13]. Compared with other baseline correction methods, the Polyfit-RSM can be the most effective method to reduce noise. Haibing et al. proposed a piecewise polynomial fitting algorithm to divide the spectral data and fit the appropriate order separately. The iterative optimization method is used to eliminate discontinuities between piecewise points [15].

An improved baseline correction algorithm is proposed based on the least-square method. [16]–[19]. Zhimin et al. proposed an adaptive iteratively reweighted penalized least-squares (airPLS). The baseline is estimated based on iteratively changing the weight of the sum of squared error computed from the fitted baseline and the original signal [16]. Jiangtao et al. used the similarity between multiple spectra to estimate the baseline and proposed an asymmetric least-square method for multi-spectral limit correction [17]. Shixuan et al. proposed an improved asymmetric least-square (IAsLS) method for baseline correction of Raman spectroscopy. The raw spectrum baseline initiates the baseline correction algorithm, and this baseline can be estimated using a polynomial fitting method. [18]. Sungjune et al. proposed a weighting scheme based on generalized logic functions called asymmetric weighted least squares smoothing for baseline correction [19]. This method iteratively estimates the noise level and adjusts the weight accordingly.

Feng et al. proposed an automatic baseline correction method based on penalty least-square method. The algorithm first linearly spreads the end of the spectrum signal and then adds Gaussian peaks to the extended range. The whole spectrum is then corrected by the adaptive smoothness parameter penalty least-square (asPLS) method [20]. Yunnan et al. proposed an iterative smoothing spline (ISREA) method with error adjustment. ISREA is simple, fast, and can generate a consistent and accurate baseline that maintains all meaningful Raman peaks [21]. Haoran et al. proposed a sparse Bayesian learning (SBL) framework for joint pure spectrum fitting and baseline correction. The results on both the simulated and the actual data sets show that SBL has a significant performance improvement [22]. Wang et al. proposed a baseline correction method based on a search algorithm (SA). The algorithm compresses the original spectral samples into a data set with a small number of data points. It then converts the peak removal process to solve artificial intelligence (AI) search. The problem is to minimize the objective function by deleting peak data points [23].

## 1.1 Research fields of Raman spectroscopy

Since the Raman spectrum of cells and other substances is not affected by water, accurate fingerprint information can be obtained in such a measurement environment. Therefore, Raman spectroscopy has achieved great success in the field of biomedicine. Jaena et al. combined surface-enhanced Raman scattering (SERS) and statistical pattern analysis to prove the label-free and susceptible classification method of exosomes which are cancer markers. This technique was claimed for new real-time diagnosis and classification of lung cancer [40]. Prochazka et al. proposed a chemometric classification method for six kinds of bacteria based on laser-induced breakdown spectroscopy and Raman spectroscopy. Principal component analysis (PCA) was used to visualize bacterial strain data, and Kohonen's self-organizing map (SOM) was used to classify bacteria [41]. Lwan et al. overcame cell heterogeneity by obtaining integrated Raman spectra covering most cells and classifying cells by support vector machines (SVM). They compared the cell spectrum, image spectrum, and integrated Raman spectrum [42]. Wonil et al. used RI-insensitive nano-laminated SERS substrates to realize label-free Raman spectrum analysis and classification of live cancer cells, with a prediction accuracy rate of 96% [43].

Therefore, Raman spectroscopy plays an essential role in the field of chemical and biochemical analysis. Zhou et al. used Raman spectroscopy to analyze the chemical structure changes of water at the critical point of fluid mixtures [28]. Yao et al. demonstrated full Raman images of individual vibrational modes at the angstrom level for a single Mg-porphine molecule, revealing distinct characteristics and proposing a new methodology for structural determination called "scanning Raman picoscopy" [29]. The deconvolution and resolution of overlapping bands in the Raman spectra of a set of coals studied by curve fitting methods have improved our understanding of natural mature coals' main structural changes [30]. He et al. used Fourier transform infrared spectroscopy (FTIR) and Raman spectroscopy to comprehensively characterize chemical structures of coals, ranging from lignite to anthracite. It was discovered from this technique that the size of aromatic clusters and graphite crystallite continuously increase, and the microcrystalline structure in coal is gradually perfected, closer to graphite structure. [31].

Raman spectroscopy plays an important role in the field of food safety. In our daily life, food is an essential source of energy, and the quality of things is related to life, health and safety. However, Raman spectroscopy can quickly identify the quality of food. Roberto et al. provided an investigation of the current proliferation of portable Raman instruments in the SERS detection of food contaminants: a summary of studies on several analytes (mainly toxins, viruses, bacteria, pesticides, contraindicated food dyes and preservatives), and reported limited detection and final use in combination with concentration or separation technology [32]. Jingjing et al. researched and explored the first full-mode liquid microextraction technology combined with SERS. It has been successfully applied to chromium speciation analysis in food and environmental matrices [33]. Jianglong et al. proposed strategy to improve qualitative sensitivity by forming effective hot spots and quantitative analysis through dilution. It can quickly detect trace amounts of tropane alkaloids (TAs) in various foods in case of emergency first response but also could be easily extended to other SERS-based analysis [34]. Santosh et al. established a system to simultaneously detect Sudan dye's adulteration and Congo red dye in red pepper powder. The adulteration of benzoyl peroxide and alloxazine in wheat flour was tested at six different concentrations of 0.05-1%. [35].

Raman spectroscopy applications are not limited to the fields mentioned above. It includes planetary exploration, agricultural and forensic applications. Because of the low cost, high efficiency, anti-interference, and ability to rapidly identify target molecules , its applications will have extensive prospects.

## 2 Overview of Raman Spectroscopy Preprocessing Methods and Classification Methods

The Raman spectra of the substance contain molecular fingerprints of different material components. When processing these Raman spectra, the baseline needs to be removed with a correction algorithm. To extract the spectral feature of Raman spectroscopy, a feature extraction algorithm, such as chemometrics, is needed. Classification of mixture Raman spectra has always been a challenging problem. At present, machine learning (ML) algorithms and artificial neural networks gain more interest for extracting Raman spectral features and classifying specimens.

### 2.1 Overview of Chemometrics preprocessing methods

Chemometrics aims to optimize the chemical measurement process and experimental design through the theories and methods of computer science, mathematics and statistics, and extract valuable information from the chemical measurement data to the maximum extent. Compared with the classical chemical and physical separation, stoichiometry can best separate the complex mixture information represented by the analysis object from the mathematical dimension. It can extract and summarize the key information closely related to the connotation and essential law object. The research of chemometrics covers the whole process of chemical measurement, which provides new ideas and methods to solve various chemistry branches. Many chemistry fields have made significant contributions to this method, such as analytical chemistry, pharmaceutical chemistry, food chemistry, environmental chemistry, etc [36].

### 2.1.1  Principal component analysis (PCA)

In Raman spectrum analysis, it is usually necessary to observe the Raman intensity corresponding to the Raman shift. With a large number of spectrum data sets, spectrum features can be identified. Raman intensity strictly corresponds to Raman shift, which provides data richness and increases data collection workload. In many cases, there is a one-to-one correspondence between Raman shift and Raman intensity, which increases the complexity of the problem analysis. If only one of the variables is analyzed, the information cannot be fully utilized. Therefore, blind reduction of indicators will lose much useful information, thus producing wrong conclusions. Data dimensionality reduction is a method to deal with high-dimensional feature data. Dimensionality reduction aims to preserve high latitude essential features, remove noise and unimportant features, and improve the data processing speed. In Raman spectrum data processing, dimensionality reduction is within a specific range of information loss, saving time and cost. Dimensionality reduction has become widely used in preprocessing method.

  PCA is the most widely used data dimension reduction algorithm [11], [37]. The main idea is to map n-dimensional features to k-dimensional features, which are all new orthogonal features known as principal components, which are reconstructed from the original n-dimensional features. The work of PCA is to find a set of mutually orthogonal coordinate axes from the original space. The selection of new coordinate axes is closely related to the data itself. In this way, the maximum variance of the first axis and the second axis of the data is orthogonal to the original axis. By analogy, such n axes can be obtained [38]–[40]. In fact, by calculating the covariance matrix of the data matrix, the eigenvalue and eigenvector of the covariance matrix are obtained. The covariance matrix is composed of the K features with the largest eigenvalue (i.e., the largest variance) is selected. In this way, the data matrix can be transformed into a new space to reduce data features [41]. Figure 1 describes the process of PCA analyzing the components of the Raman spectrum of food substance.

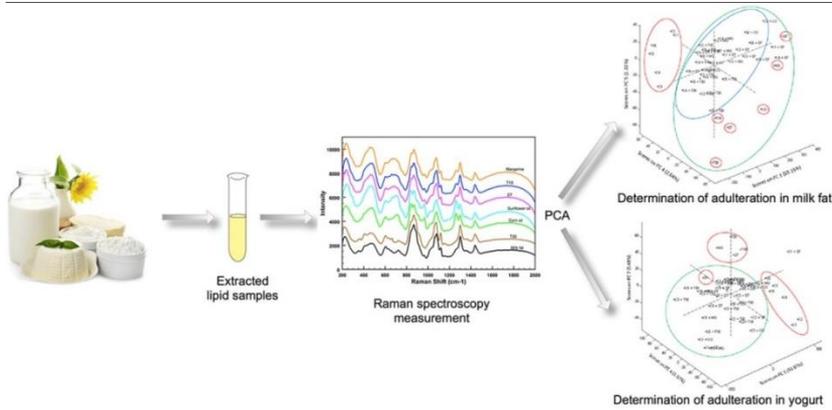

Figure 1. The process of PCA analysis of Raman spectra [42].

### 2.1.2 Cluster analysis (CA)

Cluster analysis (CA) is a method of multivariate statistical analysis of samples or indicators. It discusses a large number of samples and requires reasonable classification according to their characteristics. There is no model for reference or follow. That is, it is carried out without prior knowledge. The clustering effect depends on two factors: (1) the method of measuring distance and (2) the clustering algorithm. The method of calculating the distance between data is generally selected according to the type of data. The types of data may include numerical variables, binary variables, category variables and ordered variables. Secondly, the clustering algorithm includes K-means clustering, hierarchical clustering, clustering according to density, and clustering according to network [43]–[45]. These methods are useful for analyzing the characteristics of data.

The traditional clustering algorithm has successfully solved the clustering problem of low dimensional data. However, due to the complexity of practical application data, the existing algorithms often fail to work, especially for high-dimensional data and extensive data. Traditional clustering methods mainly encounter two problems in clustering high-level data sets. (1) There are many irrelevant attributes in high-dimensional data sets, which makes the possibility of clusters in the dimension near zero. (2) The data in high-dimensional space are sparse, and the distance between data is almost equal. However, the traditional clustering method is based on distance. Therefore it is impossible to construct clusters based on distance in high-dimensional space [46], [47]. High dimensional clustering analysis has become an important research direction of clustering analysis. The progress of random technology makes data collection more comfortable and more manageable, resulting in the database becoming larger and larger, and the analysis difficulty is increasing [48]. Consequently, high dimensional clustering is a very active field in clustering analysis and challenging work.

## 2.2 Overview of supervised learning classification methods

### 2.2.1 Support vector machine (SVM)

SVM is a binary classification model. Its basic model is defined as a linear classifier with the most considerable interval in the feature space. Its learning strategy is the maximum interval, which can be transformed into the solution of a convex quadratic programming problem. The training method of SVM is to search a hyperplane that can separate the marked training data from the hyperplane boundary. These hyperplanes are usually constructed by analyzing the data points most likely to be misclassified (that is, the data points near the candidate hyperplanes). In the classification of Raman spectra, simple Raman spectra have been classified by this method. To perform a more complex classification, SVM can also use kernel techniques to effectively map this kind of non-linearly separable original input into high feature space. The kernel techniques allow SVM to find the optimal separation hyperplane in higher dimensions. The core of SVM is to maximize the classification margin, that is, to optimize the support vector (the distance from the point to the hyperplane).

Bakhtiaridoost et al. proposed a method including Raman spectroscopy and Wavelet transform and used SVM to distinguish different types of white blood cells, and circulating tumor cells derived from breast cancer (MCF7) [49]. Jian et al. applied the particle swarm optimization to optimize SVM model's penalty and kernel parameters. It was reported that the classification accuracy of two androgens (testosterone propionate and nandrolone) in four groups of chicken samples was improved [50].

### 2.2.2 Naive Bayes (NB)

In machine learning, the naive Bayes (NB) classifier is a simple probability classifier trained by a series of strong independence of assumed features based on the Bayes theorem. It is highly scalable; hence it needs several linearly related parameters to the variables (features, predictors) for learning. The maximum likelihood estimation is the core of parameter estimation of NB. The maximum likelihood training can be completed by evaluating a closed-form expression, which only takes time on linear fitting and does not need to spend time in parameter iteration [51]. Prior probability of each class can be calculated by assuming all kinds of equal probabilities or estimated by the number of occurrences of various samples in the training set. The classifier model will assign the class labels represented by features to the problem instances. These class labels are taken from a limited set. Therefore, the NB classifier considers that the probability of each label is independent.

### 2.2.3 K-nearest neighbor (KNN)

K-nearest neighbor (KNN) algorithm is a primary classification and regression method. For a new input instance, given a training data set, find K instances nearest to the instance in the training data set. Most of these K instances belong to a particular class, and then the input instances are classified into this class. The method is based on the influence of the minority obeying the majority. Therefore, the selection of K value is significant. If the selected K value is small, the model will become complicated, and other training examples adjacent to K instances will become the model noise. Therefore, the model is prone to overfitting. If the selected K value is immense, the model will predict the data in a large field. At this time, the training instance far away from the instance will also play a role in the prediction, which is the prediction error. Therefore, the increase of K value makes the model simple. Euclidean distance is defined as the method of measuring the nearest neighbor. When the Euclidean distance is optimized,

KNN will normalize the features to prevent overfitting.

### 2.2.4 Random forest (RF)

Random forest (RF) classification algorithm belongs to a supervised learning method called decision tree. RF is a random way to build a forest composed of multiple independent decision trees. In creating each decision tree, the data is randomly sampled at first, and then a decision tree is established by completely splitting the sampled data. A new example is then presented to the RF model, in which each decision tree will rank the examples according to its selected characteristics. Finally, the output of each decision tree is merged to get the final prediction of the model. Each decision tree selects part of the training examples and randomly selects the features, which avoids overfitting to a certain extent. The model has a specific anti-noise performance and stable performance. The disadvantages of RF are mainly related to its complexity. With the increase of decision trees, the data features are refined and the forest structure becomes more massive. Hence model parameters require a machine with high computing power.

### 2.2.5 Artificial neural networks (ANN)

Artificial neural network (ANN) is a kind of machine learning technique. With the emergence and development of deep learning, ANN model's performance has been significantly improved, such as VGG, Google net, ResNet [52]. A classical neuron is equivalent to a basic processing unit, which simulates how biological neurons transmit signals, and transmits the received external stimulus signals to each neuron. The signals are acted by many neurons layer by layer, and finally, the output signals are generated at one end of the neuron to feedback the external input signals. The ANN calculates the weighted average value of all input signals and then uses a nonlinear function to generate the output. These neurons are grouped or connected to form a neural network. The unidirectional process of signal from the input layer, through one or more hidden layers, and transmitted from the output layer is called forward propagation. This kind of network structure is called feedforward neural network [53].

Figure 2. introduces a simple feedforward neural network. The purpose of neural network training is to make the parameters close to the real features possible to minimize the error between the predicted and the actual output results. Increasing the depth of the network is an excellent way to reduce the prediction error. With the deepening of the network layer, abstract representation of each layer for the previous layer is more in-depth. The target is distinguished by extracting abstract features for better classification. The fitting performance of the function increases with the network layer number increases, leading to increase the parameters. The neural network is the method of simulating the real mapping function between feature targets. More parameters means that the function can be more complex and can have more capacity to fit related features. However, the tendency of overfitting is inevitable. Hence much attention should be paid to (1) the neural network structure, (2) Selection of hyperparameters. and (3) the independent data sets for verifying the model.

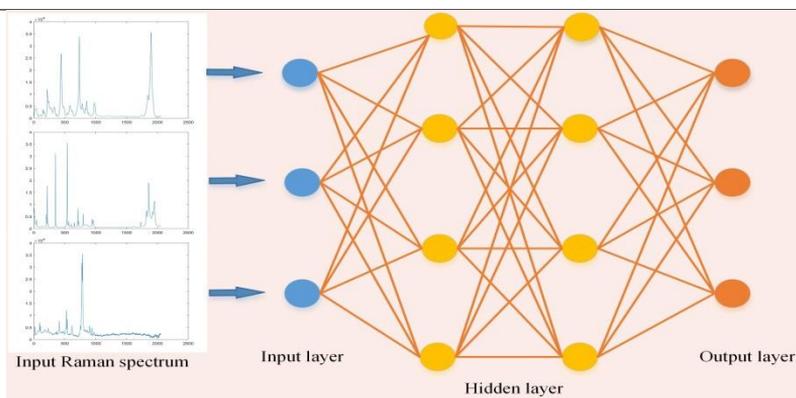

Figure 2. Workflow of a simple feedforward neural network. One-dimensional Raman spectrum is passed through the input layer, local perception and feature extraction are done in the hidden layer, and judgments are made from the output layer.

In addition to simple feedforward neural networks, various neural networks have also been applied to analyze Raman spectra mixtures. Compared with traditional perception, Convolutional Neural Network (CNN) extracts low-level features from the input, following with local perception, parameter sharing, and spatial self-sampling [54], [55]. The cooperation between different network layers makes the model more specific and sensitive. CNN was achieved to distinguish the Raman spectra of human and animal blood [56]. The proposed network consists of two preprocessing layers (noise reduction and baseline correction layer) and a fully connected classifier [57]. By optimizing the network structure, Won Bum et al. proposed a method based on a single-layer multiple-kernel-based convolutional neural network (SLMK-CNN) as an analytical tool for biological Raman spectroscopy [58]. Another method named unified CNN was proposed, which promotes Raman spectroscopy application in various fields. In the medical field, hepatitis B diagnosis using Raman spectra classification based on PCA and multi-scale CNN was proposed [59]. Based on the unique vibration characteristics between healthy cells and cancer cells, CNNs model has been used to classify tongue squamous cancer cells. It demonstrates the potential of CNNs to evaluate edges of resection sites during surgery [60]. At present, a variety of open Raman spectroscopy databases were available for developing CNN techniques and their applications in several other fields.

## 3 Application

### 3.1 Chemicals

As a supplementary method of infrared spectroscopy, Raman spectroscopy has a wide range of applications in various fields due to its fast speed, high sensitivity, and low detection dosage [61]. Raman spectroscopy is suitable for the vibration of non-polar bonds of the same atom, such as C-C, S-S, N-N bonds, etc. Symmetrical skeleton vibrations provide information observed by Raman spectroscopy. In substance analysis, Adsorption of dimethyldodecylamine oxide (DDAO) and its mixtures with Triton X-100 (TX-100) at the hydrophilic silica/water interface has been studied using total internal reflection (TIR) Raman spectroscopy and target factor analysis (TFA). [62]. When studying the influence of pressure and environment on the Raman spectrum of methane, carbon monoxide, carbon dioxide and hydrogen, a method was proposed to predict the positions of these molecules peaks [63]. In recent years, Raman spectroscopy has been proved to qualitatively and quantitatively analyze xylene isomers and their mixtures [64]. Since the Raman spectra peak corresponding to different functional groups is not the same, the experiment developed a new mixture analysis method based on nonnegative elastic net (NN-EN). The experimental results show that NN-EN can identify the compounds in the mixture with high accuracy and estimate the relative concentration with a small deviation. Besides, when the spectra of certain compounds are highly correlated, NN-EN is more stable than nonnegative lasso (NN-LASSO) [65].

As a chemical analysis tool, Raman spectroscopy was applied to study the crystallization of calcium carbonate in salt solutions at different temperatures. Using advanced mixture analysis algorithms based on Bayesian theory (BPSS), the Raman spectra of three pure anhydrous polymorphs (vaterite, aragonite, and calcite) can be recovered [66]. A new hierarchical Bayesian model was demonstrated, suitable for the nonnegative sum and one-to-one constraints related to the linear mixture's source and mixing coefficient [67]. A new hierarchical Bayesian model is studied, suitable for nonnegative sum and one-to-one constraints related to the source and mixing coefficient of linear mixtures. The performance of the proposed algorithm is evaluated through the simulation results of synthetic mixture data. It has been successfully used in the processing of Raman spectra of chemical mixtures. The experimental results show that Bayesian modeling and calculation play a key role in the quantitative study of Raman spectroscopy.

With Raman spectroscopy, SERS has become an emerging analytical technique with high speed and sensitivity. Combining it with other technologies can overcome Raman spectroscopy limitations and enhance SERS ability for detection and characterization. It has been successful in many fields [68]. SERS is a useful tool for analyzing surface adsorption phenomena. Using a sensitive SERS substrate, the adsorption isotherms of chemical substances with different binding energies were characterized [69]. SERS also realized the determination of competitive adsorption isotherms for multi-compound solutions for the first time. Chemical pollutants have always been a threat to water quality. It has scientifically proven the importance and potential of SERS technology in detecting pollutants in aqueous samples [70]. It includes the monitoring of organic and inorganic pollutants and has achieved remarkable results.

### 3.2 Food

As consumer health awareness increases, information on food ingredients needs to be more precise. However, some edible oils in the market are shoddy and adulterated. Based on the low-field nuclear magnetic resonance (NMR) spectrum characterization of six vegetable oils and binary mixtures of olive oil and three seed oils (corn, soybean and sunflower oil), the authentication and classification model was established using SVM. With the suspicious range of 10%–30 %, the SVM model provided acceptable classification results [71]. Recently, it was proved that partial least squared discriminant analysis (PLS-DA) and support vector machine classification (SVM-C) can classify olive oil samples and other vegetable edible oils with 100% and 92% accuracy rates, respectively [36]. The combination of Raman spectroscopy technology and multivariate statistical technology can be used to measure the purity and concentration of extra virgin olive oil (EVOO). Partial least squares (PLS) calibration models with 10-fold cross validation were constructed for binary, ternary and quaternary oil mixtures to determine the purity of spiked EVOO [72]. A new method for the rapid processing of Raman spectra using machine learning algorithms was developed and successfully tested for edible oil certification [73]. Therefore, Raman spectroscopy has great potential in reducing oil adulteration behavior.

Through Raman spectroscopy, we can analyze the molecular structure of meat products and the relationship between groups and obtain the concentration and distribution of animal protein, fat and other components [74]. Meat quality can be qualified using Raman spectroscopy and the stoichiometric method. Red meat from white meat on the content of horse meat on different meat days can be distinguished [75], [76]. Combining the PLS-DA and SVM classification model with Raman spectroscopy is an effective red meat detection technique [77]. In the Raman spectra analysis of the meat mixtures, the intensity ratios of 937/1003, 879/1003, 856/1003, 829/1003, and 480/1003 cm-1 spectrum pairs determined the intensities of the reference intensity of 1003cm-1 spectrum [78]. Some researchers use Raman spectroscopy to analyze edible oil produced by food manufacturers. The asymmetric least-squares smoothing (ALSS) method is used to classify edible oils, according to the Raman intensity corresponding to the unsaturated fatty acids in different animal oils to prevent the mixed sale of vegetable oil and animal oil to consumers [79]. A parameter prediction model was established based on a good fit between the spectral parameters and the chemical content. A two-stage algorithm based on Bayesian modeling and calculation was recently proposed and applied for Raman spectroscopy to quantitatively analyze the analyte concentration in complex mixtures [80]. Secondly, this method is also applied to the estimation of glucose concentration in the biopharmaceutical process. Raman spectral imaging and self-modeling mixture analysis (SMA) have been used to identify the three components mixed into complex food powder mixtures. This method can be used to simultaneously identify different components and estimate their concentrations for identification or quantitative inspection [81].

## 3.3 Medicine

The composition and quantitative analysis of drugs are essential for medical evaluation and treatment of poisoning. Because Raman spectroscopy can provide detailed chemical fingerprints, it has received more and more attention as an analytical technique in the pharmaceutical field [82]. Identifying and detecting drugs in poisoning cases usually requires time-consuming and complex chromatography techniques [83]. The use of partial least squares regression (PLSR) applied to Raman spectroscopy can quantitatively analyze the cocaine in the ternary mixture. The results show that rapid medical intervention and determination of death cause can be carried out from the quantitative cocaine toxicity through the ternary mixture analysis. The identification of fentanyl, a primary culprit in opioid overdose deaths, has become critical. When the intensity of Raman scattering is fragile, use surface-enhanced Raman spectroscopy for forensic analysis and quantification of fentanyl and use a calibration model evaluation certificate [84]. However, solvent microextraction (SME) combined with SERS is used to quickly and accurately detect illegal drug additives in health products. Through SME-SERS high-throughput and experimental testing, the level of illegal drug additives incorporated into health products is 0.1μg mg-1. The results show that this quick and easy method can effectively separate and sensitively detect illegal additives in complex samples [85]. The handheld Raman spectrometer helped Raman spectroscopy to become popular. In the fight against drug counterfeiting, a handheld Raman spectrometer was used to successfully analyze 33 of 39 product series and verify that it was 100% correct or otherwise [86]. The results show that Raman handheld devices are a reliable tool to combat drug counterfeiting.

Researchers used the subtle differences in the vibration modes of biological agents to develop a decision-making algorithm based on PLS-DA. These decision-making algorithms use spontaneous Raman spectroscopy and unlabeled plasmon-enhanced Raman spectroscopy to provide excellent discrimination capability. The experiment uses FTIR, NIR and Raman spectroscopy combined with PLSR to develop a simple on-site method to quantify the modafinil content in generic drugs [87]. Today, with the rapid development of AI, the combination of Raman spectroscopy and deep learning is a trend in the identification of drug mixtures. It has been proved that the unlabeled Raman spectroscopy technology combining DL and nonnegative least squares (NNLS) can quickly select lactose-dominated drug (LDD) formulations [88]. Protein drugs or macromolecular drugs are innovative products used to treat various diseases (such as hepatitis or cancer). Raman spectroscopy and microscopy techniques can be applied to analyze counterfeit protein-based drugs quickly [89].

### 3.4 Biological bacteria

Raman spectroscopy can analyze samples in aqueous solutions; hence it is widely used for cell mixture detection and disease diagnosis. SERS can provide multiple imaging permeability and molecular phenotypes of bladder tissue. To improve the detection and resection of non-muscle invasive bladder cancer, we quantified the ability of surface-enhanced Raman nanoparticles and endoscopic systems to classify bladder tissue as normal or cancerous [90]. Recently, microfluidic Raman biochip detection of exosomes is considered one of the most promising tools for diagnosing prostate cancer. EpCAM-functionalized Raman-active polymeric nanomaterials (Raman beads) allow rapid analysis of exosome samples within 1 hour based on a quantitative signal at 2230 cm−1 [91]. After SERS measured the plasma of nasopharyngeal carcinoma (NPC) patients and healthy volunteers, the effect of NPC radiotherapy was evaluated [92]. Raman spectroscopy combined with multivariate data analysis (such as PCA, partial minimum variance analysis, and SVM) methods creates a reliable and rapid technique for diagnosing the bacteria that cause the symptoms of urinary tract infections. The unique micro-Raman spectral features exhibited by the five bacterial biochemical components of urinary tract infections (UTIs) were successfully detected [93].

In the field of food safety, rapid and sensitive detection of pathogens is required. Based on SERS, a simple, fast, and label-free SERS mapping method was proposed to detect Salmonella enterica and Escherichia coli on silver dendrites based on PCA for analyzing a mixture of pathogens [94]. Raman spectroscopy combined with CNN was applied to identify the coryneform bacteria at the species level. The technique achieved 97.2% accuracy for all 18 types of corynebacteria. A fully connected artificial neural network (ANN) applied for Raman spectroscopy was proposed to determine the actual ratio of specific Arcobacter species in the bacterial mixture (biomass from 5% to 100%) with a regression coefficient > 0.99 [95]. SERS has shown its advantages in small molecule detection. A volume-enhanced Raman scattering (VERS) substrate was developed, which is composed of hollow nanocones at the bottom of the micro bowls (HNCMB) and uses HNCMB VERS. Substrate proved that the Raman signal of a single virus is greatly improved [96].

Table 1. In recent years, Raman spectroscopy has been used to detect mixtures for real evaluation.

| Category | Sample(s) and/or Analyte(s) | Raman and/or SERS | Pre-process method(s) | Method(s) Used | Ref. |
|---|---|---|---|---|---|
| Chemicals | Dodecyl ammonium oxide | Raman | | TFA | [62] |
| | CO, $CO_2$, $CH_4$ | Raman | | | [31] |
| | Xylene Isomers | Raman | | | [64] |
| | Liquid, powder mixture | Raman | | NN-EN | [65] |
| | Calcium carbonate | SERS | | NB | [66] |
| | Three pure anhydrous polymorphs | Raman | | NB | [67] |
| | Methanol | SERS | | | [69] |
| | Organic and inorganic pollutants | SERS | | | [70] |
| Food | Vegetable oil | Raman | | SVM | [71] |
| | Vegetable oil | Raman | | PLS-DA, SVM-C | [36] |
| | Vegetable oil | Raman | | PLS | [72] |
| | Vegetable oil | Raman | | PCA，KNN | [73] |
| | Muscle food | Raman | | Genetic algorithms | [75] |
| | Horse meat | Raman | | PCA,LDA,SDKNN | [76] |
| | Beef, venison and lamb | Raman | | PLSDA, SVM | [77] |
| | Horse meat and beef | Raman | | LDA | [78] |
| | Lard | Raman | | ALSS | [79] |
| | Glucose | Raman | | RJMCMC | [80] |
| | Non-dairy creamer | Raman | | SMA | [81] |
| Medicine | Cocaine | Raman | | PLS | [83] |
| | fentanyl | SERS | | peak-height calibration | [84] |
| | Pharmaceutical additives | SERS | | SME | [85] |
| | Solid medicine | Raman | | | [86] |
| | modafinil medicines | Raman | | PLSR | [87] |
| | lactose dominated drug | Raman | | DL，NNLS | [88] |
| | Protein drugs | Raman | | microscope | [89] |
| Biological bacteria | Bladder cancer tissue | SERS | | molecular imaging | [90] |
| | Prostate tumor | Raman | | Microfluidic Raman Biochip | [91] |
| | Serum | SERS | | PCA-LDA | [92] |
| | UTI bacteria | Raman | | PLS-DA,SVM | [93] |
| | Salmonella enteritidis and | SERS | | PCA | [94] |

| | Escherichia coli | | | | |
| --- | --- | --- | --- | --- | --- |
| | Campylobacter and Helicobacter pylori | Raman | | ANN | [95] |
| | Virus | SERS | | PCA | [96] |

# 4  Prospect

Compared with other chemical analysis and infrared spectroscopy, Raman spectroscopy has several advantages. (1) It can be applied to various water-based solvents, especially the measurement for aqueous solution. (2) It can test sample in a package (glass or thin plastic). (3) It is a rapid and nondestuctive test method, and (4) accurate molecular information can be obtained. The disadvantage is that the laser light source may damage the sample. Organic molecules can easily convert the absorbed photons into fluorescent molecules to produce a fluorescent effect, which will interfere the Raman spectrum related to molecular fingerprint. Raman spectroscopy is generally not suitable for the determination of fluorescent samples, so the reduction of fluorescence is a critical step in the study of Raman spectroscopy. Standard methods to reduce the fluorescence effect include curve fitting, filtering and denoising. At present, many baseline correction algorithms have been proposed. Secondly, the excitation wavelength can be adjusted according to different samples.

Raman spectroscopy has been widely used in the field of mixture identification as a fast and accurate method of identifying the components of substances. We need to prepare enough data for different mixture composition analysis tasks and perform modeling and analysis before testing. However, in actual experiments, only part of the Raman spectra of the mixture can be collected, limiting Raman spectroscopy application in different fields. Online storage and analysis of Raman spectroscopy data is a direction of future development. Raman spectroscopy data set sharing facilitates real-time detection and analysis of mixtures, such as RRUFF. With the continuous development of Raman spectrometer equipment, the instrument can stabilize the laser power using the integrated software. Hence resolution of Raman spectroscopy and sensitivity can be improved. With the software, automatic baseline correction and online Raman analysis can be achieved. This will speed up the application and promotion of Raman spectroscopy in the field of mixture composition analysis.

In the future, Raman spectroscopy needs to be combined with chemometrics and AI. More in-depth explorations are needed in model optimization and information fusion. The development of deep neural networks provides an excellent opportunity for the application of Raman spectroscopy. Secondly, infrared spectroscopy can be integrated into instrument optimization. Some new techniques, such as SERS, can be combined with infrared spectroscopy to enhance substance identifying capability. In the experiments, increasing the generality and coverage of the model and applying Raman spectroscopy to online detection equipment can improve substance analysis range and accuracy. At present, many researches mainly focus on the combination of Raman spectroscopy and other techniques to improve the sensitivity of substance analysis.

**Declaration of competing interest**

The authors declare that they have no known competing financial interests or personal relationships that could have influenced the work reported in this paper.

## Acknowledge

This research was financially supported by Science, Reserach and Innovation Promotion Fund (Grant No: 1383848) and National Research Council of Thailand (Grant No. 32302).